# Visualization in teaching and learning mathematics in elementary, secondary and higher education

Branko J. Malešević[1]   Ivana V. Jovović[1]   Bojan D. Banjac[1,2]

**Abstract:** In this paper we present our experience in using visualization in mathematics education. The experience with our university courses: "Computer tools in matematics" and "Symbolic algebra" provides the basis for mathematics teacher education program http://vizuelizacija.etf.rs/. The program is intended for elementary and high school teachers. The education program deals with modern techniques of visualization by using technologies such as GeoGegebra, JAVA and HTML.

## 1. INTRODUCTION

In the following sections we present our experiences in using visualization in mathematics education. In each section the concrete mathematical topics are selected and visualization techniques are briefly illustrated using GeoGebra and Java applets for different levels of education.

## 2. COMPUTER TOOLS IN MATHEMATICS

Undergraduate studies in Electrical and Computer Engineering at School of Electrical Engineering, University of Belgrade, offer elective course "Computer tools in mathematics" (http://raum.etf.rs). Students can choose this course on Computer Science and Informatics module in the fifth semester. On the other modules there is, in the third semester, a similar accompanying course "Computer practicum in Mathematics 3" of the mandatory course "Mathematics 3" (consisting of Calculus 3 – Multiple Integrals, Complex Analysis, Laplace and Fourier transformation).

"Computer practicum in Mathematics 3", among the others software, uses free software Sage. The goal of the Sage project is to create a free open source alternative to Maple, Mathematica, and MatLAB, which are the most popular non-free closed source mathematical software systems.

Students are expected to produce a final project for the course "Computer tools in mathematics" using free software Sage and GeoGebra. Each of the projects deals with a specific topic presented as a website. GeoGebra enables students to make easily HTML pages with GeoGebra applets. Now, we will present one of the applets from the course. Applet is concerned with an application of mathematics in electrical engineering.

We will introduce the ideal mass-spring-damper system with the amplitude gradually decreasing to zero and with mass $m$, spring constant $k$ and damping coefficient $b$. For the oscillatory force it holds

$$F_s = -k\, y(x), \qquad (1)$$

and for the damping force we have

$$F_d = -b\, y'(x). \qquad (2)$$

By Newton's Law of motion the total force on the body is

$$F_t = m\, y''(x), \qquad (3)$$

where $y''(x)$ is the acceleration of the mass, $y'(x)$ is the speed of the mass and $y(x)$ is the displacement of the mass relative to a fixed point of reference. Since

$$F_t = F_s + F_d \qquad (4)$$

we obtain the second order homogenous differential equation with constant coefficients of the form

$$m\, y''(x) + b\, y'(x) + k\, y(x) = 0. \qquad (5)$$

We are interested only in the case where the characteristic polynomial

$$m\lambda^2 + b\lambda + k = 0 \quad (6)$$

has complex roots. Hence, the general solution of the motion of the mass-spring-damper system is given by

$$y(x) = e^{-\frac{b}{2m}x}\left(C_1 \cos(\omega x) + C_2 \sin(\omega x)\right), \quad (7)$$

where $\omega = \frac{\sqrt{4mk - b^2}}{2m}$. The solution can be rewritten in the form

$$y(x) = A e^{-\frac{b}{2m}x} \sin(\omega x + \varphi), \quad (8)$$

where $A = \sqrt{C_1^2 + C_2^2}$ and $\varphi = \arctan\frac{C_1}{C_2}$. So the solution $y(x)$ oscillate between the two curves $y(x) = |A|e^{-\frac{b}{2m}x}$ and $y(x) = |A|e^{\frac{b}{2m}x}$. If we consider the differential equation (5) with initial conditions

$$y(0) = y_0,$$
$$y'(0) = y_1, \quad (9)$$

we obtain

$$A = 2\sqrt{\frac{m(ky_0^2 + by_o y_1 + my_1^2)}{4mk - b^2}},$$
$$\varphi = \arctan\left(\frac{y_0 \sqrt{4mk - b^2}}{by_0 + 2my_1}\right). \quad (10)$$

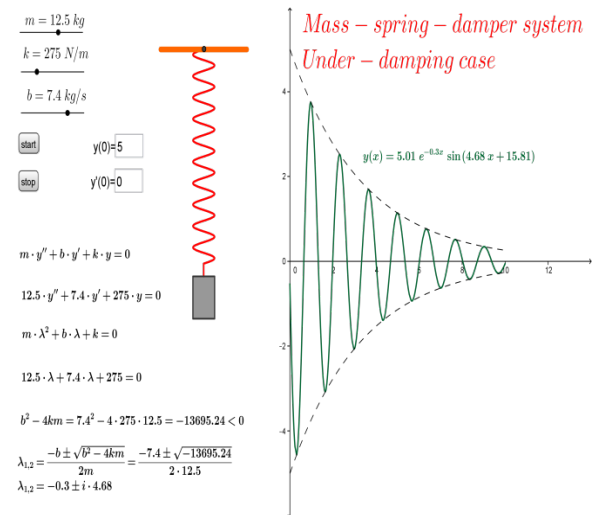

Fig. 1. Mass-spring-damper system

Now, we will give the detailed explanation of the GeoGebra applet presented on the Fig. 1. Firstly, we create four sliders for mass *m*, spring constant *k*, damping coefficient *b* and time *t*. Slider for time (not visible in Fig. 1.) is necessary for the animation. Insert Buttons "Start" and "Stop" are used to begin and terminate animation. These tools are used to execute StartAnimation[*t*] and StartAnimation[false] GeoGebra script commands with a single click on buttons. The Input Boxes " $y(0) = $ " and " $y'(0) = $ " are used to change interactively initial conditions and they are connected to object " $y0$ " and " $y0'$ ". Numbers *c* and $\beta$ are defined as $c = 4mk - b^2$ and $\beta = -\frac{b}{2m}$. We need these numbers for the written text at the bottom of the applet on Fig. 1.

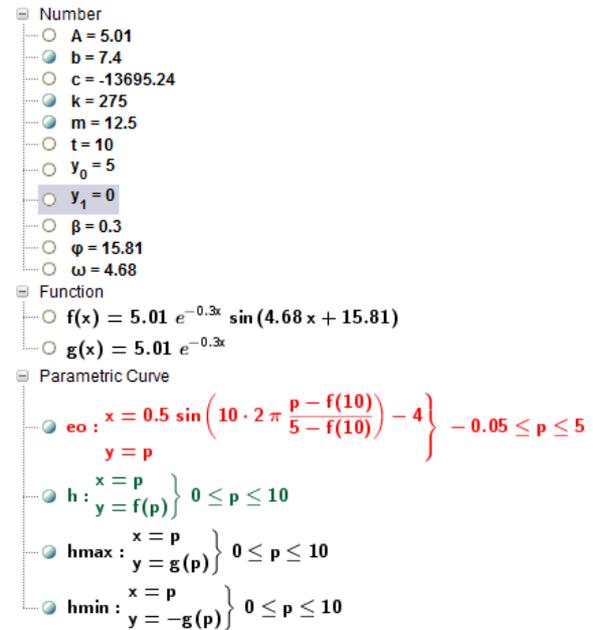

Fig. 2. Algebra view

Parametric Curve " $eo(p)$ " is used to visualize the oscillating string. The Curves " $h(p)$ "," $h\max(p)$ ", and " $h\min(p)$ " present the solution of differential equation and the curves between that solution oscillates. The rest of the applet deals with printing text with interactive objects.

## 3. SYMBOLIC ALGEBRA

Master studies in Electrical and Computer Engineering at School of Electrical Engineering, University of Belgrade, offer elective course "Symbolic algebra" (http://simba.etf.rs) on Applied Mathematics module. The course is dedicated to the study of the theory of Gröbner bases and their applications in the various fields of electrical engineering. Here we present two examples of the wide range of applications (in Robotic, Cryptography, Coding theory, ...) of Buchberger's algorithm. The algorithm is a method of transforming a given set of generators for a polynomial ideal into a Gröbner basis with respect to some monomial order. Gröbner basis of the ideal obtained in this manner is not unique. To obtain unique reduced Gröbner basis we use algorithm Tab. 3. Application has been developed in Java and it was a part of the master thesis of the Phd student Bojan Banjac, who works as a student assistant at the Department of Applied Mathematics. Talented students who choose course "Symbolic algebra" are encouraged and challenged to participate with new applications of the algorithm in conferences in this field. In the past two years, three groups of students have successfully presented their papers at the Symposium "Mathematics and Applications", Faculty of Mathematics, University of Belgrade, Serbia.

| Multivariate division algorithm [1] |
|---|
| *Input:* $f_1, \ldots, f_k, f$ |
| *Output:* $a_1, \ldots, a_k, r$ |
| $a_1 := 0, \ldots, a_k := 0, r := 0$ |
| $p := f$ |
| WHILE $p \neq 0$ DO |
|   $i := 1$ |
|   *divisionoccured* := *false* |
|   WHILE $i \leq k$ and *divisionoccured* = *false* DO |
|     IF $LT(f_i)$ divides $LT(p)$ THEN |
|       $a_i := a_i + LT(p)/LT(f_i)$ |
|       $p := p - LT(p)/LT(f_i) \cdot f_i$ |
|       *divisionoccured* := *true* |
|     ELSE |
|       $i := i + 1$ |
|   IF *divisionoccured* = *false* THEN |
|     $r := r + LT(p)$ |
|     $p := p - LT(p)$ |
| (STOP) |

Tab. 1. Multivariate division algorithm

| Buchberger's algorithm [2] |
|---|
| *Input:* $F = (f_1,...,f_k)$ *generating set for ideal I* |
| *Output:* $G = (g_1,...,g_s)$ Gröbner basis *for ideal I* |
| $G := F$ |
| $M := \{(f_i, f_j) \mid f_i, f_j \in G \land f_i = f_j\}$ |
|     WHILE $M \neq 0$ DO |
|       $(p,q) := $ *a ordered pair from M* |
|       $M := M \setminus \{(p.q)\}$ |
|       $S := S(p,q)$ |
|       $h := normal(S,G)$ |
|       IF $h \neq 0$ THEN |
|         $M := M \cup \{(g,h) \mid g \in G\}$ |
|         $G := G \cup \{h\}$ |
| (STOP) |

Tab. 2. Buchberger's algorithm

| Algorithm for obtaining a reduced Gröbner basis[3] |
|---|
| Input: $G = \{g_1,...,g_s\}$ |
| Output: $\hat{G} = \{\hat{g}_1,...,\hat{g}_k\}$ |
| $\hat{G} := G$ |
| FOR ALL $g \in \hat{G}$ DO |
|   IF $(\exists \mu \in \hat{G} \mid \mu \neq g) lt(\mu) \mid lt(g)$ THEN |
|     $\hat{G} := \hat{G} \setminus \{g\}$ |
|   ELSE |
|     $g := rem(g, \hat{G} \setminus \{g\})$ |
| FOR ALL $g \in \hat{G}$ DO |
|   $g := \dfrac{g}{lc(g)}$ |
| (STOP) |

Tab. 3. Algorithm for obtaining a reduced Gröbner basis

One of application created by students under the supervision of professor Malešević is "Vrba". The idea was to show expansion of the monomial ideal using Buchberger's algorithm and to describe in detail the way it works. Two and three dimensional graphs are used to demonstrate functionality of the algorithm for polynomials in two or three unknowns. Programming language Java was chosen, so the application can be available for all students independently from a computer they access. A combination of standard libraries for graphic user interface which are part of java execute environment and JOGL library are used for this purpose. JOGL library is created as wrapper around the OpenGL library written in program language C. Therefore, the application is not portable in the way that the most Java applications are. But this problem is solved by skilfully written files for running the applet.

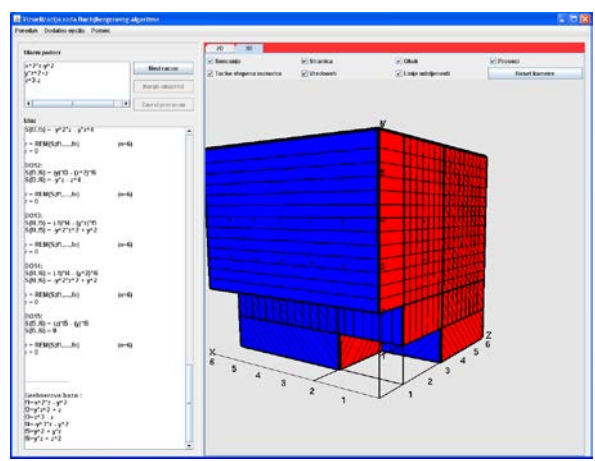

Fig. 3. Main screen of application "Vrba"

SymPy library is used for the calculation of Gröbner basis. This library requires interpreter for programming language Python. Although various Java libraries make this possible, their size combined with the low speed make application slower for launching and execution from predicted one. In order to eliminate these complications, in the later versions of the application, new specialized library for calculation is written and implemented.

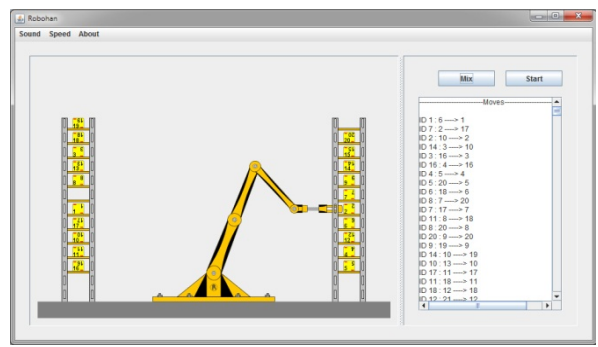

Fig. 4. Main screen of application "Robohan"

Here we will present one more application named "Robohan". It is an application of Buchberger's algorithm in Inverse Kinematics. Problem of robotic arm movement can essentially be reduced to solving system of polynomial equations using the theory of Gröbner bases. The idea for multimedia project

"Robohan" is to present in an appealing and attractive way usage of the previous algorithms. Knowledge of Computer Graphics, theory of Gröbner bases and of generating sounds was needed. As with the previous application, programming language Java was chosen, so application can be placed on the course site. Standard libraries for graphic user interface were used. Some of calculations were carried before applications starts, and some were implemented during its execution. This decision was made because the execution time was very long for some calculations. The sort algorithm was used for sorting boxes by robotic arm. One of idea was that the pre-recorded sound accompanied application. Problem was in the delay of sound reproduction. To avoid this, project participants used JSyn library to form sound waves that follows execution of application.

## 4. PROGRAM OF EDUCATION FOR TEACHERS OF MATHEMATICS

Department of Applied Mathematics, at School of Electrical Engineering, University of Belgrade, has started program of education "The visual representation of some mathematical content using computers" (**http://vizuelizacija.etf.rs/**) in 2011/12. Institute for the promotion and teaching resources Serbia (http://www.zuov.gov.rs/) accredited this program for 2013/14**.** school year. The program is designed for mathematics teachers in elementary and secondary schools. Program is conducted by members of Department of Applied Mathematics in the premises of the Computer Center of School of Electrical Engineering and in Regional Centers for Professional Development of Education Personnel.

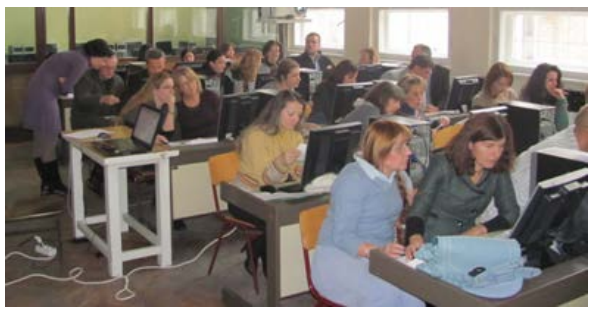

Fig. 5. Picture from the seminar

Program lasts for two days, total time is 16 hours. The training has two parts: plenary lectures and workshops. Plenary lectures focus on some interesting topics and parts of program GeoGebra related to visualization of some mathematical content using computers. During workshops, educators and participants work together to design simple applets in GeoGebra. Participants engage in an iterative process of applet creation that can be part of their teaching techniques. Here we provide a simple applet for visual representation of translation.

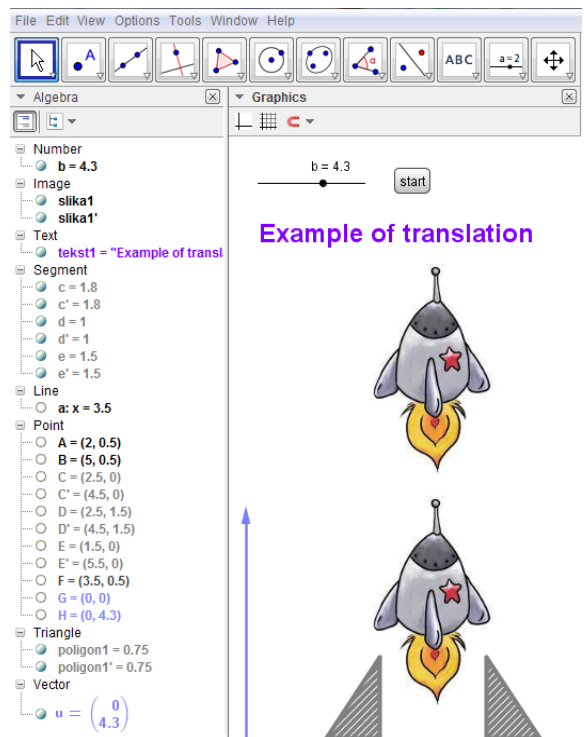

Tab. 6. Example of translation